\documentclass[12pt]{article}
\begin{document}
\title{Evidence for a Photon Mass}
\author{B.G. Sidharth\\
International Institute for Applicable Mathematics \& Information Sciences\\
Hyderabad (India) \& Udine (Italy)\\
B.M. Birla Science Centre, Adarsh Nagar, Hyderabad - 500 063
(India)}
\date{}
\maketitle
\begin{abstract}
The author's work over the past years has indicated that the photon
has a small mass $\sim 10^{-33}eV$. Recent observations from three
different viewpoints -- the time lag in cosmic gamma rays with
different frequencies, the observation of the spectra of blazars and
an analysis of the CMB power supression from the WMAP data -- all
vindicate this conclusion and remarkably, the same value.
\end{abstract}
\section{Photon Mass}
As is well known the concept of the photon grew out of the work of
Planck and Einstein, though its earliest origin was in Newton's
Corpuscular Theory. Thereafter the photon got integrated into
twentieth century physics, be it Classical or Quantum. Though it is
considered to be a massless particle of spin 1 and 2 helicity states
(as proved later for
any massless particle with spin by Wigner), it is interesting to note that there had been different
dissenting views with the photon being endowed with a small mass.\\
An apparent objection to this view has been that a photon mass would
be incompatible with Special Relativity. However it is interesting
to note that nowhere in twentieth century physics has it been proved
that the photon indeed has no mass \cite{deser}. Furthermore we are
lead to the fact that there is a minimum mass in the universe from
two different points of view. Firstly from Heisenberg's Uncertainty
Principle we have
\begin{equation}
mc^2 \cdot T \sim \hbar\label{A}
\end{equation}
If $T$ is the age of the universe $\sim 10^{17}sec$, then (\ref{A})
gives
\begin{equation}
m \sim 10^{-65}gm = 10^{-33}eV\label{B}
\end{equation}
From the author's Planck oscillator underpinning theory too we
arrive at exactly the minimum mass given in (\ref{B})
\cite{bgsfpl05}.\\
There is another simple way of arriving at (\ref{B}). We model the
dark energy by a background electromagnetic field which is an
infinite collection of independent oscillators, with amplitudes
$X_1,X_2$ etc. The probability for the various oscillators to have
amplitudes $X_1, X_2$ and so on is the product of individual
oscillator amplitudes:
$$\psi (X_1,X_2,\cdots ) = exp [-(X^2_1 + X^2_2 + \cdots)]$$
wherein there would be a suitable normalization factor. This
expression gives the probability amplitude $\psi$ for a
configuration $B (x,y,z)$ of the magnetic field that is described by
the Fourier coefficients $X_1,X_2,\cdots$ or directly in terms of
the magnetic field configuration itself by, as we saw,
$$\psi (B(x,y,z)) = P exp \left(-\int \int \frac{\bf{B}(x_1)\cdot \bf{B}(x_2)}{16\pi^3\hbar cr^2_{12}} d^3x_1 d^3x_2\right).$$
$P$ being a normalization factor. At this stage, we are thinking in
terms of energy without differentiation, that is, without
considering Electromagnetism or Gravitation etc as separate. Let us
consider a configuration where the magnetic field is everywhere zero
except in a region of dimension $l$, where it is of the order of
$\sim \Delta B$. The probability amplitude for this configuration
would be proportional to
$$\exp [-((\Delta B)^2 l^4/\hbar c)]$$
So the energy of \index{fluctuation}fluctuation in a region of
length $l$ is given by finally as is well known, the density
(Cf.ref.\cite{tduniv}),
$$B^2 \sim \frac{\hbar c}{l^4}$$
So the energy content in a region of volume $l^3$ is given by
\begin{equation}
\beta^2 \sim \hbar c/l\label{4e1}
\end{equation}
This energy is minimum when $l$ in (\ref{4e1}) is maximum. Let us
take $l$ to be the radius of the Universe $\sim 10^{28}cms$. The
minimum energy residue of the background dark energy now turns out
to be exactly the value in (\ref{B}), which as we will see has
indeed experimental
confirmation.\\
While it would be tempting to identify the photon mass with
(\ref{B}), it would be correct to say that there are a number of
experimental upper limits to the mass of the photon as we will see.
These limits have become more and more precise \cite{lakes,it}. The
best limit so far is given by
\begin{equation}
m_\gamma < 10^{-57}gm\label{4eb1}
\end{equation}
that is, the photon mass would be very small indeed!\\
\section{Theoretical Support}
The author has argued from different points of view to show that the
photon has a mass given by (\ref{B}) \cite{tduniv,bhtd,notefpl}. We
now touch upon yet another
theoretical indicator \cite{bgsafdb}.\\
We first observe that as is well known \cite{newmanjmath}, Maxwell's
equations can be written in the following form
\begin{equation}
{\bf \Psi} = \vec{E} + \imath \vec{B},\label{4eb3}
\end{equation}
\begin{equation}
\vec{\nabla} \times {\bf \Psi} = \imath \dot{{\bf \Psi}}\label{4eb4}
\end{equation}
\begin{equation}
\vec{\nabla} \cdot {\bf \Psi} = 0\label{4eb5}
\end{equation}
Equations (\ref{4eb3}) to (\ref{4eb5}) will be useful in the sequel.\\
We next observe that Maxwell's equations have been deduced in a
fashion very similar to the Dirac equation, from first principles
\cite{gersten}. In this deduction, we use the usual energy momentum
relation for the photon
$$E^2 - p^2 c^2 = 0$$
and introduce matrices given by
$$S_x = \left(\begin{array}{ll}
0 \quad 0 \quad 0\\
0 \quad 0 \quad -\imath\\
0 \quad \imath \quad 0
\end{array}\right)\, , S_y = \left(\begin{array}{ll}
0 \quad 0 \quad \imath\\
0 \quad 0 \quad 0\\
-\imath \quad 0 \quad 0
\end{array}\right)\, ,$$
\begin{equation}
S_z = \left(\begin{array}{ll}
0 \quad -\imath \quad 0\\
\imath \quad 0 \quad 0\\
0 \quad 0 \quad 0
\end{array}\right)\, , I^{(3)} = \left(\begin{array}{ll}
1 \quad 0 \quad 0\\
0 \quad 1 \quad 0\\
0 \quad 0 \quad 1
\end{array}\right)\, ,\label{4eb6}
\end{equation}
from which we get
$$\left(\frac{E^2}{c^2} - {\bf p^2}\right) {\bf \Psi} = \left(\frac{E}{c}I^{(3)} + {\bf p \cdot S}\right){\bf \Psi}$$
\begin{equation}
- \left(\begin{array}{ll}
p_x\\
p_y\\
p_z
\end{array}\right) \, ({\bf p \cdot \Psi}) = 0,\label{4eb7}
\end{equation}
where $\Psi$ is a three component wave function and in general bold
letters denote vector
quantities.\\
Equation (\ref{4eb7}) implies
\begin{equation}
\left(\frac{E}{c} I^{(3)} + {\bf p \cdot S}\right) {\bf \Psi} =
0,\label{4eb8}
\end{equation}
\begin{equation}
{\bf p \cdot \Psi} = 0,\label{4eb9}
\end{equation}
where $S$ is given in (\ref{4eb6}). There is also an equation for
${\bf \Psi^*}$ namely
\begin{equation}
\left(\frac{E}{c}I^{(3)} - {\bf p \cdot S}\right) {\bf \Psi^*} =
0,\label{4eb10}
\end{equation}
\begin{equation}
{\bf p \cdot \Psi^*} = 0,\label{4eb11}
\end{equation}
It is then easy to verify (Cf.ref.\cite{gersten}) that with the
substitution of the usual Quantum Mechanical energy momentum
operators, we recover equations (\ref{4eb3}) to (\ref{4eb5})
for ${\bf \Psi}$ and its complex conjugate.\\
Recently a similar analysis has lead to the same conclusion. In fact
it has been shown that under a Lorentz boost
\cite{dvg,ignat,tucker},
\begin{equation}
\left(\begin{array}{ll}
\Psi'\\
\Psi^{*'}\end{array}\right) = \left(\begin{array}{ll}
1 - \frac{({\bf S \cdot p})}{mc} + \frac{({\bf S \cdot p})^2}{m(E+mc^2)} \quad \quad 0\\
0 \quad \quad \quad \quad 1 + \frac{({\bf S \cdot p})}{mc} +
\frac{({\bf S \cdot p})^2}{m(E+mc^2)}\end{array}\right)
\left(\begin{array}{ll}
\Psi\\
\Psi^* \end{array}\right)\label{4eb12}
\end{equation}
We would like to point out that equations (\ref{4eb4}),
(\ref{4eb5}), (\ref{4eb8}) to (\ref{4eb12}) display the symmetry
$${\bf p} \to -{\bf p} \quad , \Psi \to \Psi^*$$
We now invoke the Weinberg-Tucker-Hammer formalism
(Cf.\cite{tucker}) which gives, for a Lorentz boost, the equations
\begin{equation}
\phi'_R = \left\{ 1 + \frac{{\bf S \cdot p}}{m} + {({\bf S \cdot
p})^2}{m(E + m)}\right\} \phi_R,\label{4eb13}
\end{equation}
\begin{equation}
\phi'_L = \left\{ 1 - \frac{{\bf S \cdot p}}{m} + {({\bf S \cdot
p})^2}{m(E + m)}\right\} \phi_L,\label{4eb14}
\end{equation}
where the subscripts $R$ and $L$ refer to the states of opposite
helicity, that is left and
right polarised light in our case.\\
We observe that equations (\ref{4eb12}) and
(\ref{4eb13})-(\ref{4eb14}) are identical, but there is a curious
feature in both of these, that is that the photon of
electromagnetism is now seen to have a mass $m$.
\section{The Experimental Scene}
(i) Nevertheless there are experimental tests, in addition to those
mentioned above, which are doable. It is well known that for a
massive vector field interacting with a magnetic dipole of moment
${\bf M}$, for example the earth itself, we would have with the
usual notation (Cf.ref.\cite{it})
$${\bf A}(x) = \frac{\imath}{2} \int \frac{d^3 k}{(2\pi )^3}{\bf M \times k} \frac{e^{\imath k, x}}{{\bf k}^2 + \mu^2} =
- {\bf M \times \nabla} \left(\frac{e^{-\mu r}}{8\pi r}\right)$$
\begin{equation}
{\bf B} = \frac{e^{-\mu r}}{8 \pi r^3} | {\bf M}| \left\{\left[
\hat{r} (\hat{r} \cdot \hat{z}) - \frac{1}{3} \hat{z}\right] (\mu^2
r^2 + 3\mu r + 3) - \frac{2}{3} \hat{z} \mu^2
r^2\right\}\label{4ec4}
\end{equation}
Considerations like this have yielded as noted in the past an upper
limit for the photon mass, for instance $10^{-48}gms$ and
$10^{-57}gms$. Nevertheless (\ref{4ec4}) can be used for a precise
determination of the photon mass.\\
(ii) With a non zero photon mass we would have, for radiation
(Cf.ref.\cite{tduniv})
\begin{equation}
E = h \nu = m_\nu c^2 [1 - v^2_\gamma /c^2]^{-1/2}\label{4ec1}
\end{equation}
From (\ref{4ec1}) one would have a dispersive group velocity for
waves of frequency $\nu$ given by (Cf. also ref.\cite{vigierieee})
\begin{equation}
v_\gamma = c \left[1 - \frac{m^2_\gamma c^4}{h^2
\nu^2}\right]^{1/2}\label{4ec2}
\end{equation}
We would like to point out that (\ref{4ec2}) indicates that higher
frequency radiation has a velocity greater than lower frequency
radiation. This is a very subtle and minute effect and is best
tested in for example, the observation of high energy gamma rays,
which we receive from deep outer space. It is quite remarkable that
we may already have witnessed this effect-- higher frequency
components of gamma rays in cosmic rays do indeed seem to reach
earlier than their lower frequency counterparts \cite{pav}.\\
(iii) Another test for the massive photon comes from the
observations of De Angeles and co-workers of the MAGIC Telescope
team. They have observed an inexplicably large transparency to gamma
rays, by using the Imaging Atmospheric Cherenkov Telescope and
MAGIC. Their conclusion is that this anomalous observation can be
reconciled with standard blazar emission models provided the photon
oscillates to a Light Axion like Particle (ALP) in extra galactic
magnetic fields. The ALP again has the very same mass of
$10^{-65}gms (10^{-33}eV)$. These considerations have been
successfully applied to the Blazar
$3C 279$ \cite{angelis}.\\
(iv) A further confirmation for exactly the same photon mass comes
from the observation of a small residual energy at the edge of the
universe, recently \cite{mercini,bgsfpl}: There has been a wealth of
data from the WMAP. One of the intriguing findings is that the dark
energy domination and the CMB power supression, both occur around
the same red shift and energy scale - corresponding to the energy
scale of the Hubble radius $\sim 10^{-33}eV$ exactly the same
minimum energy as before in (\ref{B}).
\section{Remarks}
1. De Broglie himself \cite{debroglievig} believed that the photon
has a mass, a view shared by a few others as well. Interestingly in
this context in 1940 and 1942, De Broglie published two volumes on
the Theory of Light, La mecanique ondulatoire du photon Une nouvelle
theorie de la lumiere, the first volume, La lumiere dans le vide
(Paris, Hermann, 1940); the second volume, Les interactions entre
les photons et la
matiere (Paris, Hermann, 1942) \cite{debroglie1,debroglie2}.\\
2. Laboratory diffraction experiments several years ago indicated a
photon mass similar to (\ref{B}), showing that the vacuum is a
dissipative medium \cite{vigierieee}.\\
3. In fact if we start with the Langevin equation in a viscous
medium then as the viscosity becomes vanishingly small, it turns out
that the Brownian particle moves according to Newton's first law,
that is with a constant velocity. Moreover this constant velocity is
given by, for any mass $m$,
\begin{equation}
\langle v^2 \rangle = \frac{kT}{m}\label{4e1a}
\end{equation}
We would like to study the case where $m \to 0$. Then so too should
$T$ for a meaningful limit. More realistically, let us consider
(\ref{4e1a}) with minimal values of $T$ and $m$, in the real world.
We consider in the Beckenstein-Hawking Black Hole temperature
formula the entire Universe so that the mass $M$ is $\sim
10^{55}gms$. The justification for this is that the Universe mimics
a black hole, as shown in detail several years ago by the author.
This can be seen in a simple way by the fact that the size of the
Universe is given by the Schwarzchild radius:
$$R \approx \frac{2GM}{c^2}$$
Further the time taken by light to reach the boundary at a distance
$R$ from a given point is the same as for a black hole of the mass
of the Universe. Substitution gives
\begin{equation}
T = \frac{10^4}{10^{32}} \sim 10^{-28}K\label{4e3a}
\end{equation}
We next consider in (\ref{4e1a}),m to be the smallest possible mass
encountered earlier, viz.,
\begin{equation}
m \sim 10^{-65} gms\label{4e4a}
\end{equation}
Equation (\ref{4e4a}) has been obtained as we saw from different
points of view, e.g. the Planck scale underpinning for the Universe.
Substitution of (\ref{4e3a}) and (\ref{4e4a}) in (\ref{4e1a}) gives
$$
\langle v^2 \rangle = \frac{kT}{m} = \frac{10^{-16} \times
10^{-28}}{10^{-65}} = 10^{21}, i.e.$$
\begin{equation}
v = c \, (cm/sec)\label{4e5a}
\end{equation}
We can see from (\ref{4e1a}) and (\ref{4e5a}) that the velocity $c$
is the velocity of light! So $m$ in (\ref{4e4a}) indeed represents
the mass of the photon.


\begin{thebibliography}{99}
\bibitem {deser} Deser, S. (1972). \emph{Ann Inst. Henri Poincare, Vol.XVI} (Paris, Gauthier-Villors), pp.79.
\bibitem {bgsfpl05} Sidharth, B.G. (2004). \emph{Found.Phys.Lett.} 17 (5), 2004, pp.503-506.
\bibitem {lakes} Lakes, R. (1998). \emph{Phys.Rev.Lett.} 80, (9), pp.1826ff.
\bibitem {it} Itzykson, C. and Zuber, J. (1980). \emph{Quantum-Field Theory} (Mc-Graw Hill,
New York), pp.139.
\bibitem {tduniv}  Sidharth, B.G. (2008). \emph{The Thermodynamic Universe}
(World Scientific, Singapore) (and references therein).
\bibitem {bhtd} Sidharth, B.G. (2006). \emph{Found.Phys.Lett.} \textbf{19}, 1, 2006, pp.87ff.
\bibitem {notefpl} Sidharth, B.G. (2006). \emph{Found.Phys.Lett.} 19 (4), 2006.
\bibitem {bgsafdb} Sidharth, B.G. (2009). \emph{Annales Fondation L. De Broglie} 33
(3), 2009.
\bibitem {newmanjmath}  Newman, E.T.  (1973). \emph{J.Math.Phys} \underline{14}, (1), pp.102.
\bibitem {gersten} Gersten, A. (1999) \emph{Found.Phys.Lett.} 12, (3), pp.291--298.
\bibitem {dvg} Dvoeglazov, V.V. and Gonzalez, J.L.Q. (2006). \emph{Found.Phys.Lett.} 19, (2),
pp.195ff.
\bibitem {ignat} Ignatiev, A. Yu. and Joshi, G.C. (1996). \emph{Mod.Phys.Lett.A.} 11, pp.2735--2741.
\bibitem {tucker} Tucker, R.H. and Hammer, C.L. (1978). \emph{Phys.Rev.D.}
Vol.3, No.10, 2448ff.
\bibitem {vigierieee} Vigier, J.P. (1990). \emph{IEEE Transactions of Plasma Science} 18, (1), pp.64--72.
\bibitem {pav} Pavlopoulos, T.G. (2005). \emph{Phys.Lett.B.} 625, pp.13-18.
\bibitem {angelis} Roncadelli, M., De Angelis, A., and Mansutti, O.
(2008) in \emph{Proceedings of Frontiers of Fundamental and
Computational Physics} Ninth International Symposium (Ed. B.G.
Sidharth, et al.) (AIP Conference Proceedings, Melville, 2008) 1018.
\bibitem {mercini} Mersini-Houghton, L., {\it Mod.Phys.Lett.A.}, Vol.21, No.1, (2006), 1-21.
\bibitem {bgsfpl} Sidharth, B.G. (2006). \emph{Found.Phys.Lett.} 19,
2006, pp.499-500.
\bibitem {debroglievig} De Broglie, L. and Vigier, J.P. (1972). \emph{Phys.Rev.Lett.}
\underline{28}, pp.1001--1004.
\bibitem {debroglie1} De Broglie, L. (1940). \emph{La mecanique
ondulatoire du photon Une nouvelle theorie de la lumiere} Vol.I
(Paris, Hermann).
\bibitem {debroglie2} De Broglie, L. (1942). \emph{Les interactions
entre les photons et la matiere} Vol.II (Paris, Hermann).
\end{thebibliography}
\end{document}